# The Dynamics of Human Society Evolution: An Energetics Approach

Ram C. Poudel[1, 2], Jon G. McGowan [2]

[1]Department of Mechanical Engineering, Central Campus Pulchowk, Institute of Engineering, Tribhuvan University, Nepal

[2]Department of Mechanical and Industrial Engineering, University of Massachusetts, Amherst, USA

Last updated: 3/31/2019 4:17 PM

## ABSTRACT

Human society is an open system that evolves by coupling with various known and unknown (energy) fluxes. How do these dynamics precisely unfold? Energetics may provide further insights. We expand on Navier-Stokes' approach to study non-equilibrium dynamics in a field that evolves with time. Based on the 'social field theory', an induction of the classical field theories, we define social force, social energy and Hamiltonian of an individual in a society. The equations for the evolution of an individual and society are sketched based on the time-dependent Hamiltonian that includes power dynamics. In this paper, we will demonstrate that Lotka-Volterra type equations can be derived from the Hamiltonian equation in the social field.

Keywords: energetics; social field theory; capital; capabilities; evolution.

*In memory of Richard N. Adams (1924 – 2018), the author of The Eighth Day*

## 1.0 Introduction

Human life is a complex system. It is very difficult to interpret human life based on its physical properties alone. Let's postpone the genesis of life for now; we will focus here on human life after it emerges. Whether one reads Longfellow [1] or Tesla [2], they each point out that human life is a movement. What is the nature of this movement? Instead of making an eponymous hypothesis, we conjecture that the human system and its underlying movements are not much different from many other systems around us. In order to make some sense of a conscious human being, we may need a higher level of comprehension [3] along with some generalizations beyond the classical field theories.

The classical field theories define the potential energy of an object within the field of the other object that shares the same property such as mass or charge, or (di)pole strength. A force is a gradient of the potential energy. Many phenomena in nature can be interpreted in terms of four fundamental forces: electromagnetic, gravity, strong and weak forces. Are the myriad phenomena in nature governed by just these four fundamental forces? Many of us assume such a notion to be true. Nonetheless, some scholars are in search of more general principles [4, 5, 6, 7] that may be applicable to living beings. Mark Buchanan [8] suggests looking at the patterns rather than the people in order to study human systems and behavior. Robert Laughlin [9] is in search of organizational forces beyond the microscopic rules: a physical principle where the whole can be greater or less than the sum of the parts. Such organizational force may be at play in a field generated by a dipole, electric or magnetic. The orientation of one dipole has an effect on the energy of the other dipole within this field. If these two dipoles constitute two separate systems, we may be able to see an effect similar to the organizational force, where the whole becomes not only more than but very different from the sum of the parts [10].

We proposed there exists a new type of force, especially among social beings. Earlier we have made a case for the social field theory [11] by a generalization of the classical field theories. Based on this theory, we defined Hamiltonian of an individual in society. The equation for the evolution of an individual is sketched here based on the time-dependent Hamiltonian that includes power dynamics (a forcing term). In this paper, we will expand Navier-Stokes' approach to studying the non-equilibrium dynamics in a field that evolves with time and establish that the Lotka-Volterra type equation can be derived from the time-dependent equations in the social field.

This article is organized in sections. In Section 2 we start with background information followed by a short review of the available literature on energetics. Social field theory is summarized in Section 4 and the subsequent sections present some properties of the field. With a ground-work spanning Sections 6 through 9, we propose equations of motion for an individual and society in Section 10. A quick summary of the implications for physics and economic science are documented in Section 12. We conclude following a brief discussion in Section 13.

## 2.0 Background

Measurement and logic are general methods in all sciences. In natural science, we base our measurements on concepts like space, time and energy (or force- which is its derivative). Social scientists have identified many similar forces, however, it has yet to quantify social forces and social energy. There are many examples in science where a proper quantification of the relationship among variables of interest in a mathematical language has opened up a structured reasoning that has eventually contributed to the advancement of science. The social science can't be an exception.

Is there a fundamental difference between natural and social sciences? No doubt they are two cultures [12] existing within the limits of human observation. A part of social sciences is exploratory, which revolves around the development of noble concepts or theories. Social theorization involves a richness/freedom beyond a structured line of reasoning exercised in many physical theories. One needs to cut to the core of the issues beyond human observation. Along the lines of thought by Riemann, we argue that an effort connecting these two sciences should not be hindered by too narrow views, and progress should not be obstructed by traditional prejudices such as the conservation laws or invariance principles. An open system like human society may also be inferred based on many other open systems we analyze using thermodynamic principles.

How do we communicate the traditions of knowledge and wisdom between social science and physical science? Is there a language that can cross these man-made boundaries? Based on our partial experience of natural science and encounters with colleagues across this divide, we think there are various concepts that may cross these boundaries. Energy is the simplest of all such concepts. We argue that energetics can provide a viewpoint from which the subject appears in its greatest simplicity, not necessarily the most fundamental one. Even the most complex creation of nature, a human being, must adhere to the correspondence principle and the laws of nature. It will be a misapprehension to assume that Mother Nature has a set of different laws for the human being.

### Dynamics: Kinematics and Kinetics

The dynamics of an open system, such as a human being, can be described in various approaches. These may include subjective and objective approaches utilizing some measurable properties. The subjective method may include a multiplicity of theories including the natural selection principles. Objective methods normally try to codify the dynamics utilizing the physical concepts. Some of these concepts are energy, entropy, and information. As these three concepts are mathematically related to one another [13], it does not make much difference theoretically to approach the problem using either of these threads. Following Schrodinger [14] and Prigogine [15], many physicists approach the problem utilizing entropy (or negantropy) concept. Georgescu-Roegen [16] made a seminal attempt to interpret the economic processes in terms of entropy. However, it is not easy to characterize the source term for entropy or information-based formulations for open and evolving human dynamics. The source term for energy is the power which can be perceived easily by laymen and specialists alike. Even though we may be limited in the details of dynamics, a provisional description based on energetics may provide some insights into the dynamics and evolution of human society. As noted by Ostwald, no other general concepts find application in all domains of science that include both natural and social sciences.

**Kinematics**: It is a branch of mechanics that deals with the movement of an object without reference to the underlying force, or the source term in general. In a conservative field, a kinematic description may suffice to characterize the movement if the nature of the force remains the same throughout. However, this is not the case for an open and evolving field like the social field. In an evolving social field, the force and energy may change autonomously over time. We will discuss this topic in later sections.

The father of modern economic science Adam Smith perceived an "invisible hand" back in the 18th century, however, we are yet to decipher the term. It could be a reason many theories in economic science are based on the kinematic description. It is equally possible that we authors have misunderstood/misapprehended the term. A source term like "invisible hand" is a key to understanding even the simplest concepts of supply and demand that may lead to the rationalization of the price of an object.

**Kinetics:** The dynamics of a system can also be described in the language of kinetics. Kinetics considers movement in tandem with the underlying forces or the source term in general. What is the source term in social dynamics? According to Bertrand Russell, it is power [17]: "The fundamental concept in social science is Power, in the same sense in which Energy is the fundamental concept in physics." This observation is in accordance with one of the fundamental equations governing an open system – the rate of change of energy(E) is equal to power (P), or $dE/dt = P$. Human being /society is an open system in which matter, energy, entropy, and information flow in and out of the system's boundaries.

A target of this paper is to develop provisional "equations of motion for social systems" in the way Wolfgang Weidlich [18] has long sought for. The equations we have developed for the social system are based on kinetics. The equations are energetic descriptions of the social system which takes the source term – the social power– into account. In the following section, we review briefly the science of energetics as it is relevant to this study.

## 3.0 Literature Review
### Energetics

Energy is a concept that appears in every science. The ontology of energy was not well established as we perceive it today until the handiwork of Scot William Macquorn Rankine in 1850. Rankine proposed a new science "energetics" in 1855. Energetics is the science of studying energy and its transformation. In a modern sense, it may include production, distribution, and dissipation of energy such as those that occur in a biological cell sometimes referred to as bioenergetics. No other general concept finds application in all domains of science [19]. This science got its currency on the foundation of the first law of thermodynamics, aka the principle of conservation of energy.

In the case of an open and evolving human social system, it may not be that meaningful to search equivalent concepts such as mass, inertia, momentum etc. Such an inherent challenge inhibits extension of the classical Newtonian mechanics into the social system. This challenge led curious minds like Alfred J. Lotka to rely on energetics to understand evolution [20]. In Energetics, Lotka saw a physical principle competent enough to extend our systematic knowledge to natural selection. This is unfinished business; something that never took off the ground [21], a history of which is documented briefly by Richard Adams [22]. Evolution through the lens of thermodynamics has been a topic of various studies [23, 24, 25, 26]. Prigogine advanced the thermodynamics of evolution formulation based on entropy dynamics [27]. In order to expand the concept of natural selection, Weber et al. [28] proposed a thermodynamic approach that is more appropriate to biological systems. The approach considers a biological system as embedded in the web of energetic-information relations. Recently, the Constructal Law by Adrian Bejan is gaining some momentum in order to interpret evolution both in animate and inanimate systems [29, 30].

Can energy flow alone create forms and structures? Or is structure/organization the basis of energy flowing through the system? Should such questions have a universal answer? Does the answer depend on whether we are talking about the living or non-living world [31]? Regardless of where we start from on this chicken-and-egg question, there remains a basic question: what is the physical reason that natural processes move in the direction of the origin and evolution of life?

Ronald Fox [32] argues that biological organization and evolution are a consequence of the flow of energy through matter. The motion of inanimate beads in the electric field [33] seems to evolve in time to form structures. The structure so formed also shows a self-healing mechanism. Why is there Life [34]? Contrary to the common observation, Moroitz and Smith [35] argue that the continuous generation of sources of free energy by abiotic processes may have forced life into existence and emergence. This argument may support Peter Mitchel's argument that life is a mechanism for energy transduction, in general. There could be multiple viewpoints to look at a complex adaptive system such as human society.

Whether one looks from one way or the other, energy flow is at the core of human life and evolution. So it must also be true of human society. Human society evolves and develops not only by utilizing energy that we know is in the domain of natural science, but in fact human society itself is a field of energy. Accordingly, we proposed a 'social field theory' that quantifies the energy of an individual in the social field. Regarding the structure/organization, we rely on facts created by earlier scientific research, especially the postulates by Niels Bohr [36] that energy levels are also quantized in the social field. His postulates go along with our effort to understand the first principles and causes in the Aristotelian sense. The laws of nature may have to fit together seamlessly across the living and non-living world.

### Classical Field Theory: Generalization

Let x be a parameter characterizing property such as mass or charge or the (di)pole strength of matter. According to the classical field theory, the potential energy PE of an object x at distance r in the field of object X

$$\text{PE} = -k \frac{X_i x_j}{r_{ij}^\alpha}.$$  *Equation 1*

For simplicity, assume $\alpha$ = 1. The force field F can then be expressed as

$$F = k \frac{X_i x_j}{r_{ij}^2}.$$  *Equation 2*

Table 1 summarizes the classical Gravitational field, Electrostatic field and Magnetic field in terms of generalized equations above.

Table 1: Parameter and Equation of Classical Field Theories

| Fields | Parameter | Two Objects X | x | Potential Energy (PE) | Force(F) | Constant k |
|---|---|---|---|---|---|---|
| Gravitational | Mass | M | m | $\frac{Mm}{r}$ | $k\frac{Mm}{r^2}$ | G |
| Electrostatic | Charge | Q | q | $\frac{Qq}{r}$ | $k\frac{Qq}{r^2}$ | $\frac{1}{4\pi\epsilon_o}$ |
| Magnetic | Pole strength | $m_1$ | $m_2$ | $\frac{m_1 m_2}{r}$ | $k\frac{m_1 m_2}{r^2}$ | $\frac{\mu_o}{4\pi}$ |

Parallel to the development of classical field theories [37] in natural science, many social scientists have made attempts to utilize these same concepts. Harald Mey [38] has put together such efforts in a monograph – translated from the German by Douglas Scott. However, none of these attempts has been successful so far. John Martin [39] reviewed isomorphic attempts in the realm of social science. There are more examples in spatial science [40] and economics [41]. Recently, Neil Fligstein and Doug McAdam [42] made an attempt to invigorate the field theory in social science. They interpret large-scale social dynamics by means of an interconnected "strategic action field" that anchors interaction and meaningful membership to the field.

None of these field theories is in the language of energy. Social force and social energy is yet to be quantified in theoretical sociology literature [43]. In order to quantify the social energy of an individual, we will need to characterize humans and human society first. We will review these subtle concepts in the following section.

## 4.0 Social Field Theory

There are many types of field theories in social science [44]. The Social Field Theory (SFT) that we summarize here was born out of an effort by one of us to understand the link between energy access and poverty dynamics [45]. Each of us may has a unique view of the world around us. In order to support our views of the nexus between energy and poverty, we have derived a rationale from the accepted field theories in classical mechanics. Especially, the SFT was inspired by Bohr's theory of the H-atom, which connects classical and quantum mechanics in a way many engineering students may find easy to understand.

The Earth's magnetic field deflects high energy ions making the biosphere habitable. This is a fundamental property of the blue planet earth that sets it apart from many other planets. This magnetic field may have something to do with what we do. On this epistemology, we have modeled the social field as analogous to the magnetic field. The social field is thus characterized in terms of Social Strength (S), Individual Strength (I) and the Social Distance (r). We characterize the entire spectrum of human beings at different level of experience and understanding by a single variable called the individual strength. This variable has a bearing on the idea of the pole strength in a magnetic field. The social distance, according to Wright [46], is the relation of social entities to others measuring the degree of their contact or isolation.

Our characterization of society is in accordance with sociologist Paul F. Lazarsfeld [47] who came up with the following three key societal variables:
- Global Properties: not based on information of individual members on Society;
- Analytical Properties: based on information of individual members on Society; and
- Structural Properties: based on data about the relations among members.

In line with the classical field theory, we have mapped these properties to the variables in the Social Field Theory as presented in Table 2.

Table 2: Characteristic variables of Social Field

| Lazarfeld | Social Field Theory |
|---|---|
| Global Properties | Social Strength (S) |
| Analytical Properties | Individual Strength (I) |
| Structural Properties | Social Distance (r) |

The Social Strength (S) compares the global properties; Individual Strength (I) to the analytical properties; and the Social Distance(r) to the structural properties of society, to follow Lazerfield's nomenclature.

By following Lazarsfeld, we can go around one of the complex problems in physics consisting of multi-body systems. Even characterizing the dynamics of the nucleons of He-atom is a non-trivial task. The social strength (S) represents an outcome of all such known and unknown dynamics that may exist among members of a human society. This holistic organizational parameter, to follow Joseph Needham [48], can be very different from the sum of the parts. It is our hope that this macroscale variable may be inferred based on big data that is available nowadays.

Human beings seem to be an incomplete or unfinished animal as anthropologist Clifford Geertz [49] has reiterated Nietzsche's view. *Homo sapiens* may be physically weak and incomplete but they possess an open-ended mind that can go beyond its current state. Even if all humans may be created equal, the human mind may be able to construe various dimensions of its own reality, some real and others imaginary. It may not be possible to address all such idiosyncratic dimensions that characterize a human being. To surmount our ignorance of a complex human being, we simply postulate that the individual strength (I) is a parameter in an n-dimensional space. The individuate strength describes a sum total of all dimensions that may be meaningful to an individual. Along the same vein, the social strength (S) is also an n-dimensional space. However, not all dimensions of S may matter to an individual in question.

This characterization of the social field led us to codify field theories in a language of energetics. The potential energy (PE) of an individual in the social field, is

$$\text{PE} = -k \frac{S_i I_j}{r_{ij}}. \qquad \qquad Equation\ 3$$

This equation suggests that PE of an individual is not absolute but depends on the underlying society. Equation 3 provides a conceptual framework to understand poverty [45]. Lee Smolin has suggested earlier that the property of an object is not absolute but rather depends on the environment in which we access the property. His assertion seems to us more obvious in the social field.

Here we follow natural units such that k = 1, and introduce a change of variable. We define the reciprocal of the social distance as trust vector $\Gamma$, i.e. $r \times \Gamma = 1$. Accordingly, the PE of an individual is S I $\Gamma$. A force is equal to the change of potential energy per unit distance that is sometimes known also as a potential gradient. Based on equation 3, the social force F on an individual is

$$F = \frac{S_i I_j}{r_{ij}^2} = S\,I\,\Gamma^2. \qquad \text{Equation 4}$$

The two hypotheses of the social field are [11, 45].
- HP01: Social Field is a quasi-conservative field, defined as a field for which total energy is a monotonic function of time.
- HP02: Energy levels in the social field are quantized in similar notion as in established models of an atom, Bohr's theory [36] of the hydrogen atom and Schrödinger's equation.

What are the justifications for these hypotheses? The first hypothesis [HP01] was inspired by our hands-on experience of some developing societies around the world. This hypothesis is supported in part by a correlation that exists between the Energy Development Index and the Human Development Index. Energy access is a prerequisite for human development as well as wealth creation. A recent study [50] relates the global rates of energy consumption quantitatively to a very general metric of global economic wealth. The second hypothesis [HP02] assumes continuity of living and non-living worlds. Human society at large must be governed by the same laws of nature we witness, for example, in the hydrogen atom. These two hypotheses provide a rationale to the social field in accordance with which many consequences can be deduced. In the future, some of these consequences may be compared to the facts that may come out of some empirical method. We don't yet have a clear sight of any empirical method to be utilized.

## 5.0 Properties of a Social Field

A social field is an open system that evolves with time. The evolution of parameters defining the social field makes it a non-inertial frame. In a later section, we will provide a reason why a social field consisting of living beings tends to be an autonomous system.

## Open System

In the case of an open system mass, energy (including entropy and information) can flow in and out of the system's boundaries. Human beings are an open system; so is human society. A governing equation for such a system, Figure 1, can be expressed in terms of power. Following usual notation from thermodynamics;

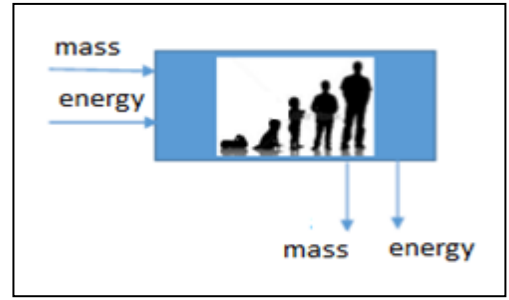

Figure 1: Open human system

$$\frac{dE_{in}}{dt} - \frac{dE_{out}}{dt} - \frac{dU}{dt} = P_{net} = P_{generation} - P_{dissipasion} \qquad \text{Equation 5}$$

Here, U is an internal energy and P$_{net}$ is net power generation internal to the system.

## Evolving Field

An evolving field is a field for which the parameters characterizing the field may change over time. Suppose the mass of earth (M) is changing with time. The gravitational field around us, quantified by Newton as $GM/r^2$, can then be considered an example of an evolving field. The social field is an evolutionary field in which variables characterizing the field can be expressed mathematically such that S = S(t), I = I(t). These variables not only change with time but also follow a gradual sequence or direction for a reason.

## Non-inertial frame

A non-inertial field is defined as a field in which acceleration may vary with time. Mathematically, $\frac{d^n x}{dt^n} \neq 0$ for n > 2; i.e. acceleration is not a constant. Let us compare Equation 4 with Newton's law of motion F(net) = ma. The acceleration term a ~ S Γ². This term is a function of time which implies that a social field, in general, is a non-inertial field.

We know from classical mechanics that a pseudo force that is a function of acceleration and other parameters comes into play in the analysis of a non-inertial frame. Many of us may have a feel for such pseudo forces in various experiences such as weightlessness in an elevator moving down with some acceleration.

## Autonomous System

An autonomous system can generate spontaneous force and a resulting motion by itself. An external force is not mandatory for such a motion. A Human being is an example of an autonomous system so is its ensemble -- a human society. A Human can utilize somatic and exosomatic energy sources to generate movements of various types. In general, an autonomous motion is a property of a living organism. The collective action of biological cells results in motions of various kinds at various scales. Likewise, the collective action of an ensemble of humans in the form of a society results in motion of various kinds, some of which may entail social evolution. A physical explanation for the autonomous motion of human society is presented in a later section.

## Energy: Hierarchical field

Hierarchy is an important concept in science. An object in a hierarchical structure can be arranged by rank in terms of its property. This concept extends from the periodic table to the food chain in the ecosystem. In a modern periodic table, elements are arranged by their atomic number. The tropic levels in the food chain are ranked in terms of relation with the primary source of energy. These levels describe how energy may cascade in an ecosystem. Hierarchical structure is a basic fact for both biotic and abiotic worlds [28].

Following HP02, we postulate that the social field is hierarchical in reference to the total energy of an individual. That is to say, human beings can be ranked in terms of the total energy in the social field. Just like many other classical fields, the total energy is composed of two forms of energy,

$$\text{Potential Energy (PE)} = -\frac{S\,I}{r}, \qquad \text{Equation 6}$$

$$\text{Kinetic Energy (KE)} = \frac{1}{2}\frac{S\,I}{r}. \qquad \text{Equation 7}$$

$$\text{Hence, total Energy (TE)} = KE + PE = -\frac{1}{2}\frac{S\,I}{r}. \qquad \text{Equation 8}$$

In the social field we equate the kinetic energy to capital $C_1$, and potential energy to the capabilities $C_2$, of an individual. This capital in the social field is a different term than its common use. Unlike the case in economic science [51, 52], the capital in the social field is a well-defined term that has the unit Joules equivalent in the natural units. The total energy in the social field is the Hamiltonian of an individual, $\mathcal{H} = \mathcal{H}(C_1, C_2, t)$. Alternatively, $\mathcal{H} = \mathcal{H}(S, I, r, t) = \mathcal{H}(S, I, \Gamma, t)$.

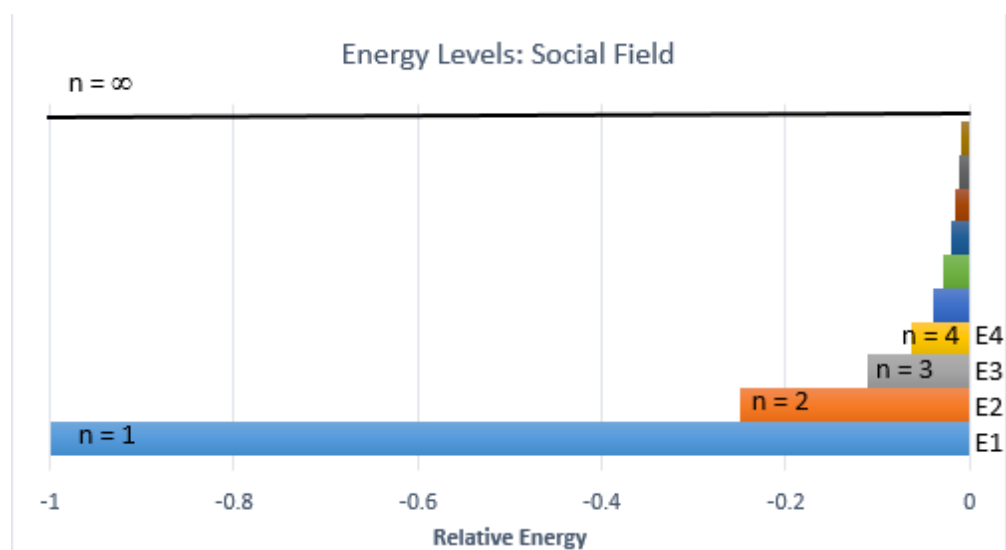

Figure 2: Relative energy levels

Figure 2 depicts the energy levels quantized in relative scales of $E_n = E_1/n^2$; n is the principle quantum number. This example is taken from H-atom. It is our postulate [HP02] that energy in the social field is also quantized in some similar fashion. A difference in energy levels $\Delta E$ is pronounced at the ground state but is not that apparent at higher quantum states.

## Entropy

Entropy is a physical quantity having multiple interpretations. Following the analogy from thermodynamics entropy can be written as

$$ds = \frac{dQ}{T} = \frac{d(SI/r)}{S/\bar{r}} = \frac{d(SI\Gamma)}{S\bar{\Gamma}}$$

Or
$$= dI\frac{\Gamma}{\Gamma} + I\frac{d\Gamma}{\Gamma} + I\frac{\Gamma}{\Gamma}\frac{dS}{S}.$$
 
Equation 9

For an open system, the source of entropy could be internal and external. Prigogine has broken entropy change ds into two components: i) the transfer of entropy across the boundary of the system, $d_eS$, and ii) the entropy produced within the system, $d_iS$. The second law dictates only about the latter, $d_iS \geq 0$. Equation 9 provides three sources of the change in entropy in the social field.

If we may ignore the change in $\Gamma$, Equation 9 reduces to s= $\bar{I} \log(SI)$. This equation is similar to the Boltzmann equation s = $\kappa$ ln W, where $\kappa$ = $R/N_A$ is the Boltzmann constant, and W is the thermodynamic probability of a macrostate.

## Social Field and Social System

A social field is a field of energy, a natural physical foundation characterizing a field. It is composed of an ensemble of units defining a society. In the case of human beings, a society may be defined as an ensemble of individuals interacting and influencing one another.

A social system, however, is composed of both a human system and a natural system. An analysis of social systems is much more complex [53] than a social field, which is beyond the scope of this paper. A human system may involve etiquettes, laws, hierarchy etc. that we have developed for the orderly functioning of human society. Inheritance, patent system, geopolitical economic boundaries may not be the domain of the natural social field we enumerate here in the language of energetics. The social scientists may provide an important insight into how we can superimpose the human system with an underlying social field to make any analysis complete and meaningful.

## 6.0 Measurement Space: Hyperspace

A phase space is a multidimensional space in which each axis corresponds to one of the coordinates required to specify the state of a physical system. A measurement space for the social field is an n-dimensional phase space of class C2; n being the total number of dimensions that matters to the society. Note that C2 has a different meaning than $C_2$. For each dimension, we have two sub-dimensions: potential energy (capabilities) and kinetic energy (capital). The class C2 represents these two sub dimensions. Figure 3 presents a measurement space for one of the dimensions of the social field. A point in the phase space corresponds to the state of a human being at a given point of time. The measurement space of the social field is thus an n-dimensional phase space of class C2 that evolves in time. We call this hyperspace $e_2^n(t)$-space for the sake of brevity.

Let O be the origin of $e_2^n(t)$-space with respect to some point of reference. The social field is an evolving field, hence the reference point also changes position in relation to some absolute scale. Along x-axis, we have capital and along y-axis we have capabilities on a normalized scale. A human being is being represented by a dot, an ordered-pair expressed in terms of capital and capabilities. The ensemble of dots enclosed by the triangle represents a society $\Omega$ under study.

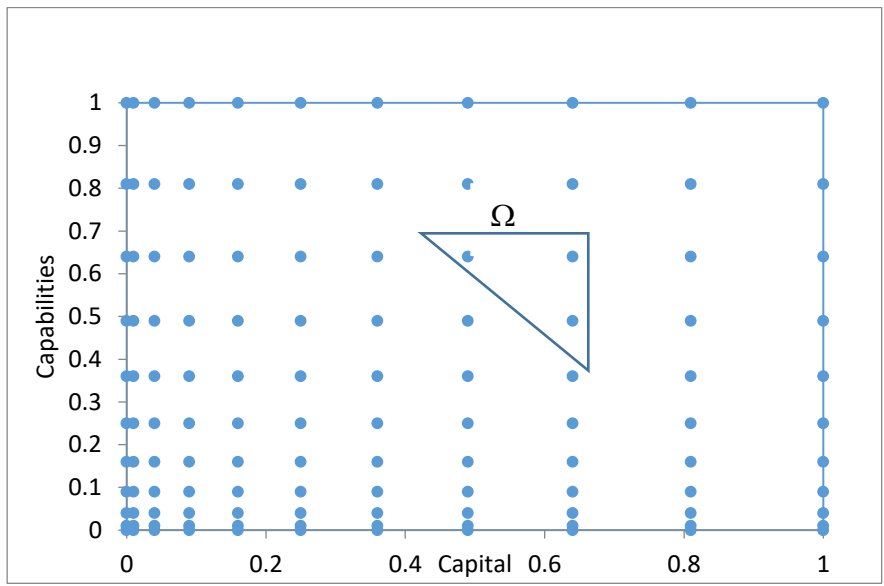

Figure 3: Measurement space for a social field

## 7.0 Social Potentials

The concept of thermodynamic potentials was introduced by Pierre Duhem in 1886. A thermodynamic potential is a scalar quantity used to characterize the thermodynamic state of a system. Thermodynamics consists of two types of variables: extensive and intensive. E (energy), s (entropy), V (volume) and N (number of particles) are extensive variables while T (temperature), p (pressure) and $\mu$ (chemical potential) are intensive variables. In thermodynamics, the functional dependence of entropy s(E,N, V) on E, N, and V is called the *fundamental equation* [54]. Equation 10 relates these variables

$$dE = Tds - pdV + \mu dN \qquad \text{Equation } 10$$

An equivalent version of the fundamental equation of social field in terms of energy is $\mathcal{H} = \mathcal{H}(s, \Gamma, N)$. The change in the Hamiltonian can be characterized by some change in s, $\Gamma$ and N. Hence, based on the multivariate calculus, we may write

$$d\mathcal{H} = \frac{\partial \mathcal{H}}{\partial s} ds + \frac{\partial \mathcal{H}}{\partial \Gamma} d\Gamma + \frac{\partial \mathcal{H}}{\partial N} dN \qquad \text{Equation 11}$$

Based on the Social Field Theory, following the footsteps of Duhem and Gibbs, we have come up with Social potentials – namely Social Temperature, Entropy Strength and Social Potential. These potentials are defined as follows

$$\left(\frac{\partial \mathcal{H}}{\partial s}\right)_{\Gamma,N} = S\Gamma = Social\ Temperature \qquad \text{Equation 12}$$

$$\left(\frac{\partial \mathcal{H}}{\partial \Gamma}\right)_{s,N} = SI = Entropy\ Strength \qquad \text{Equation 13}$$

$$\left(\frac{\partial \mathcal{H}}{\partial N}\right)_{s,\Gamma} = \sigma = Social\ Potential \qquad \text{Equation 14}$$

Expressed in terms of these social potentials, a fundamental equation in the social field thus will have a form:
$$d\mathcal{H} = S\Gamma ds + SI\ d\Gamma + \sigma dN. \qquad \text{Equation 15}$$

The Schwarz' theorem for a thermodynamic potential ϕ (xi, xj) can be written as
$$\frac{\partial}{\partial xi}\left(\frac{\partial \phi}{\partial xj}\right) = \frac{\partial}{\partial xj}\left(\frac{\partial \phi}{\partial xi}\right). \qquad \text{Equation 16}$$

Earlier, John von Neumann [55], in an attempt to extend analogies between thermodynamics and economics, had come up with function ϕ (X, Y) whose role appears to be similar to that of thermodynamic potentials in phenomenological thermodynamics. In economics, there are similar sets of equations known as Slutsky conditions [56]. The social field, however, is a non-inertial frame and interaction between individuals in the filed tends to be non-reciprocal. For these reasons, the Schwarz' theorem may not hold true in the social field. Mathematically this is the equivalent of saying the Hamiltonian $\mathcal{H}$ is an inexact differential, i.e.

$$\frac{\partial}{\partial x_i}\left(\frac{\partial \mathcal{H}}{\partial x_j}\right) \neq \frac{\partial}{\partial x_j}\left(\frac{\partial \mathcal{H}}{\partial x_i}\right). \qquad \text{Equation 17}$$

The social potentials, represented by Equations 14 - 16 are parameters relevant to the development metrology [11], and may play an important role in economic growth, development and evolutionary process. The dynamics of the evolution of human society can also be expressed in terms of these metrics. Nevertheless, we choose to adopt an approach similar to that of classical mechanics in the subsequent sections in order to keep it simple for ourselves and our readers.

## 8.0 Living vs non-living

Living beings respond to an external stimuli whereas non-living objects react. Objects react only to influences acting upon them at the instant that those influences act – a concept known in physics also as object egotism [57]. Here we define the reaction as a subset of the response. Our first approximation is that a living creature has the ability to go beyond local action-reaction symmetry described by the Newton's third law: $R_{AB} = -R_{BA}$. There may be multiple reasons for this asymmetry. Object egotism is a fundamental concept underlying Newtonian physics. We human beings are more than that. As a living creature we can scale, lead or lag behind reaction forces. Hence the general response of human being to external stimuli **R** may be expressed as:

$$R(t) = -\alpha(r)\,\mathbf{R}(t \pm \tau);\qquad\qquad\text{Equation 18}$$

The response depends on the current time, history and future projections i.e memory and foresight. Here $\alpha$ is a scale factor, and $\tau$ lead/lag time.

This demarcation of living and non-living may create more questions than it may answer. This preliminary definition, however, is inspired partly by the conventional approaches of physics that aim for tracing the phenomena of nature back to the simple laws of mechanics [58]. This definition, even if it is primitive, may resonate with many observations we encounter in the hierarchical social field. We may respond differently whether an action being imposed upon us is by our peers, subordinates or superiors in terms of power. We propose this formalism to build a base with mainstream classical physics and to move up the ladder to understand complex human life better. This approach may strengthen the wisdom of many scholars who believe in the continuity of the living and non-living world.

Consider an ensemble of human beings chosen arbitrarily from Figure 2; we may call this physical ensemble a society $\Omega$. In Figure 4, i and j are indices counting the human being Mi in the multi-body social field. G is a centroid of some kind, and R is a reaction force. In the social field, H is an inexact differential in general as defined in Equation 17.
The reaction force relates to the Hamiltonian following Equation 19 based on the Green's theorem. If the KE and PE functions of H are independent of each other, the reaction force may reduce to zero.

$$\oint R(r,t)dr = \iint \left(\frac{\partial \mathcal{H}_2}{\partial C_1} - \frac{\partial \mathcal{H}_1}{\partial C_2}\right) dC_1 dC_2 \qquad\qquad\text{Equation 19}$$

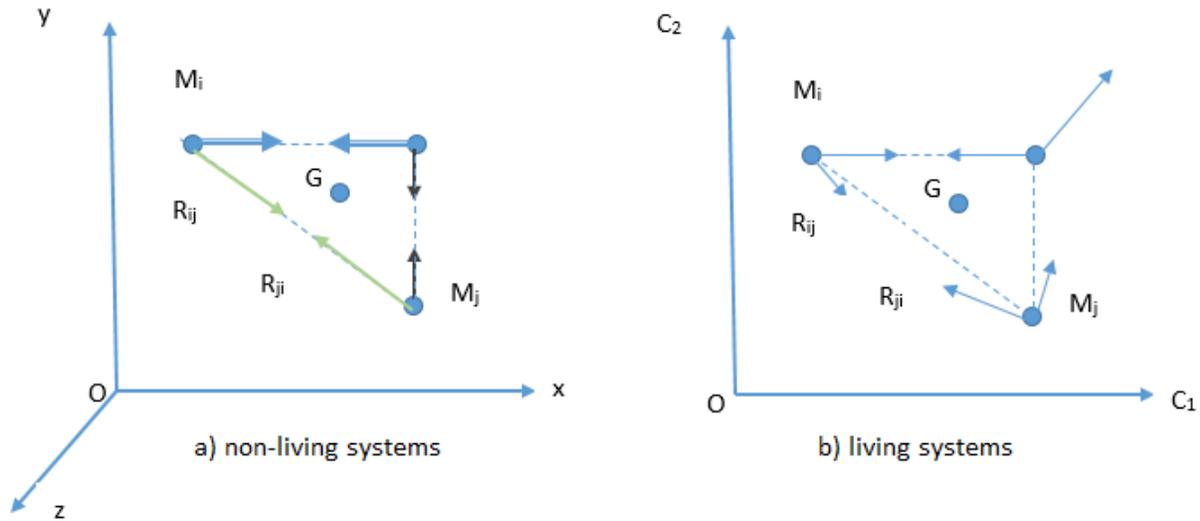

Figure 4: Non-reciprocal interaction in a social field

Figure 4 portrays an interaction of a systems of particles Mi for living and non-living systems. In many non-living systems, as shown in Figure 4(a), the internal forces $R_{ij}$ and $R_{ji}$ are equal and opposite and have same line of action. Hence, their sum $R_{ij} + R_{ji} = 0$. This is not the general case for living systems. For a living system, Figure 4(b), $R_{ij} + R_{ji} \neq 0$. The internal forces may not always be balanced between conscious human beings. Hence, a multi-body physical ensemble (society $\Omega$ as marked in Figure 3) will have a net force plus a moment around O. This net force/moment renders the social field autonomous. In addition, a living system also possesses an intrinsic force $R_{ii}$ of various degrees. Consciousness is the ability of a living creature to sense its environment plus itself. A human being can utilize its consciousness to move alone and/or with society up the energy ladder, Figure 2, by also utilizing intrinsic force $R_{ii}$. The d'Alembert's principle [59] may not hold true in the Social Field. That is to say, the system of the external forces acting on human beings Mi and the system of the effective forces of the human beings are not equipollent. The d'Alembert's principle is normally valid for a system where $R_{ij} + R_{ji} = 0$, and $R_{ii}$ is absent.

## 9.0 Autonomous System

As explained above, an ensemble of human beings can generate a force plus moment by its own mechanism, mainly because the internal forces Rij may not balance, in general, with each other. Such a system does not require an external intervention in order to change the current state of the system. This is a feature of the living world that we don't often see in the non-living world. In the case of most of the rigid body systems, the internal forces balance each other. The system reacts mostly to an external force and other induced dynamic forces. In the case of human beings, we also possess intrinsic force Rii. This force, even if at times repressible, is the very reason a social system is an autonomous system. An autonomous system is generally composed of units (sub-system) that can pump energy in and out the system. In other words, the subsystem (or constituents) of an autonomous system must consume energy in order to transform energy into work of some kind.

## 10.0 Equation of Motion

The equations of motion (EOM) describe, in general, a time evolution of a state of systems. We want to reiterate at the outset that we are proposing the EOM in the social field, not for the social system. A social system is a complex amalgam of human-made and natural systems. Hence, these EOMs would not be able to account completely for the complex dynamics we witness in human society. Nonetheless, a concept of the social field may provide a new canvas on which the complex EOMs of social systems can be sketched.

Many EOMs are based on Newton's second law, F = ma. The EOM of the human social system, however, is expressed in terms of power for a reason. In fact, the equation we propose for the social field, is a power equation Power P = F × v, where power is defined as the rate of change of energy. The change of energy is expressed in terms of the total derivative of the Hamiltonian H = H(S, I, r, t) in the framework of Navier-Stokes' equations. In short, we present an expression for the dynamics of human evolution in a language of energetics.

### Why Navier-Stokes?

The Navier-Stokes' equations are the cornerstone of fluid dynamics at various lengths and time scales. These equations belong to a class of fundamental equations of non-equilibrium statistical physics [60]. The momentum and energy equations can handle source and sink terms. These terms may represent energy generation and dissipation influencing the autonomous social field. Hence, this class of equations makes perfect sense for a quasi-conservative field [HP01].

### Equation of Motion: Individual

A society has been characterized by a set of variables: Social Strength (S), Individual Strength (I), Trust Vector ($\Gamma = 1/r$) and Time (t). Based on SFT, the Hamiltonian $\mathcal{H} = \mathcal{H}$ (S, I, r, t) corresponds to the total energy of the social field under analysis. Its time evolution can be written in total derivative form as

$$\frac{d\mathcal{H}}{dt} = \frac{\partial \mathcal{H}}{\partial t} + \frac{\partial \mathcal{H}}{\partial S}\frac{dS}{dt} + \frac{\partial \mathcal{H}}{\partial I}\frac{dI}{dt} + \frac{\partial \mathcal{H}}{\partial r}\frac{dr}{dt}.$$  Equation 20

The three macro variables (S, I and r) of society are not yet developed enough in social science in order to define and measure. Hence, the equations of the social field shall be presented here based on the Hamiltonian comprised of composite variables, capital ($C_1$) and capabilities ($C_2$) of individuals in the society as,

$$\frac{d\mathcal{H}}{dt} = \frac{\partial \mathcal{H}}{\partial t} + \frac{\partial \mathcal{H}}{\partial C_1}\frac{dC_1}{dt} + \frac{\partial \mathcal{H}}{\partial C_2}\frac{dC_2}{dt}.$$  Equation 21

The reaction forces among human beings give rise to forces of various kinds resulting ultimately to power dynamics. Following the discussion in Section 8, these forces can be grouped into extrinsic $F_{ex}$ (surface force) and intrinsic $F_{en}$ body forces. The surface force results mainly from interactions among individuals, i.e $j \neq i$, whereas the body force is intrinsic to an individual.

$$\text{Ri} = \sum_{j=1}^{n} \alpha(r) R_{ij} (t \pm \tau) \, for \, j \neq i \quad \text{Surface Force | Extrinsic}$$
$$R_{ii} \quad \text{Systematic force} \quad\quad\quad \text{Body Force | Intrinsic}$$

*Equation 22*

These forces drive the change DH/Dt of the Hamiltonian in the autonomous social field and vice versa. By including source/sink terms $\dot{\mathbb{Q}}$ to account for other generation and dissipation:

$$\underbrace{\frac{\partial \mathcal{H}}{\partial t}}_{\text{Local Derivative}} + \underbrace{\frac{\partial \mathcal{H}}{\partial C_1}\frac{dC_1}{dt} + \frac{\partial \mathcal{H}}{\partial C_2}\frac{dC_2}{dt}}_{\text{Convective Derivative}} = \underbrace{(F_{en} + F_{ex})\frac{dr}{dt} \pm \dot{\mathbb{Q}}}_{\text{Forcing or Power Terms}}.$$

*Equation 23*

In terms of the trust vector $\Gamma$, the RHS of the equation modifies as

$$\frac{\partial \mathcal{H}}{\partial t} + \frac{\partial \mathcal{H}}{\partial C_1}\frac{dC_1}{dt} + \frac{\partial \mathcal{H}}{\partial C_2}\frac{dC_2}{dt} = (F_{en} + F_{ex})\frac{1}{\Gamma^2}\frac{d\Gamma}{dt} \pm \dot{\mathbb{Q}}.$$

*Equation 24*

For many of us, we may care more about a relative change in the Hamiltonian with reference to some datum D, or the amount of change accumulated over a time window dt. The relative change $\Pi(r, t)$ may be expressed in several ways, some of which are presented below.

- $\mathcal{H}(r, t) - \mathcal{H}_D(r, t)$; D for datum.
- $\mathcal{H}_f(r, t) - \mathcal{H}_i(r, t)$
- $\frac{\mathcal{H}_f(r,t)}{\mathcal{H}_i(r,t)}$
- Log return r(log) = $\ln\left(\frac{\mathcal{H}_f(r,t)}{\mathcal{H}_i(r,t)}\right)$

*Equation 25*

The nature of the equation of motion of an individual or society would not be affected by a change of variables. Equations 23 or 24 describes the dynamics of the evolution of an individual in the social field. The following section presents equations governing the dynamics of the social field itself.

### Equation of Motion: Society

A human society is an ensemble of individuals influencing one another in the social field. Members of a society can influence each other in varieties of ways mainly in terms of energy, entropy and information as well as through an exchange of matter. Suppose there are N individuals in a society $\Omega$. The Hamiltonian Hs of a society can be aggregated as the sum total of the Hamiltonian in the ensemble. Hence,

$$\mathcal{H}_s(n, t) = \sum_{k=1}^{N} \mathcal{H}_k(n, t).$$

*Equation 26*

We can also represent $\mathcal{H}_s$ in terms of the probability distribution function in the $e_2^n(t)$-space.

$$\text{Normalize} \int \mathcal{H}(n,t)dr = 1, \, and \, \mathcal{H}(n,t) \geq 0$$

*Equation 27*

On the same basis as Equation 27, an equation of motion for a human society will have a form

$$\frac{\partial}{\partial t}\mathcal{H}_s(n,\ t) = -\sum_{i=1,2}^{n}\frac{\partial \mathcal{H}_s(n,t)}{\partial C_i}\dot{C}_i + \left(F_s(n,t) + F_b(n,t)\right)\dot{r}_n \pm \dot{\mathbb{Q}}_s \qquad \textit{Equation 28}$$

The summation on the RHS covers all dimensions and the class C2 over the $e_2^n(t)$-space. An aggregated multi-body equation in a society leads to an implicit Fokker-Planck equation. Equation 28 is a multivariate Fokker-Planck equation describing time evolution of $\mathcal{H}_s$ in the absolute scale. Following a different approach and characterization, Dirk Helbing [61] earlier introduced social field/forces on the basis of Boltzmann-Fokker-Planck equations.

### Coarse-graining: Reduced order model

What are the major dimensions that are important for a society? We depend on social scientists to help us on this coarse-graining. One of the field theorists in sociology, Pierre Buordieu has implied three major forms of capital [62]. Accordingly, we propose to contract the n-dimensional social field to $\mathbf{R}^3$ to make it manageable. These three reduced dimensions of social field are: i) economic b) cultural and c) social. Following the SFT, some logical consequences can be drawn to interpret trends in social capital brought forward by Putnam [63]. We will leave it to social scientists to evaluate if an assertion based on energetics may make sense to interpret the declining trends in social capital across many societies. In the following section we present that Lotka-Volterra type equation that can be derived from Equation 28 by simplifying the assumptions of the social field.

### Lotka-Volterra Equation

Lotka-Volterra (LV) equations are coupled first-order nonlinear equations that describe the evolution of two interdependent quantities. This simplified model was developed independently by Lotka and Volterra around the first-quarter of the 20th century. These deterministic equations are easy to explain in order to demonstrate a nonlinear dependency of a pair of variables of interest, and hence become popular as a standard example for hands-on modelling exercise in many academic disciplines. Interesting dynamical concepts such as the attractor, limit-cycles and others can be illustrated utilizing the LV equation. These equations are extended in real-world applications in various ways.

For two interdependent variables x and y, the LV equation can be expressed by
$$\frac{dx}{dt} = k_1 x - k_2 xy; \frac{dy}{dt} = -k_3 y + k_4 xy \qquad \textit{Equation 29}$$

We can multiply the first equation by y and the second by x, and add them together to get a simplified version (with $k_4 = k_2$) as,
$$x\dot{y} + y\dot{x} = (k_1 - k_3)\,xy + k_2 xy(x - y). \qquad \textit{Equation 30}$$

Next, we establish that the time-dependent Hamiltonian equation, Equation 28, reduces to LV equation, Equation 30, under special conditions. Let's assume the following form of the Hamiltonian to illustrate the concept:
$$\mathcal{H}(C_1, C_2, t) = C_1 + C_2 + k\,\frac{C_1 \cdot C_2}{C_1 + C_2} \qquad \textit{Equation 31}$$

where k = k (Δt, Ω) is a parameter that describes the symbiotic relationship between capabilities and capital [11]. The third term on the RHS of Equation 31 represents an effect of the positive (or negative) feedback loop, Figure 5, between these two variables which are interdependent in the social field Ω. If these variables are independent (coupled but with no feedback loop) this translates to k = 0; which reduces Equation 31 to the common definition of total energy.

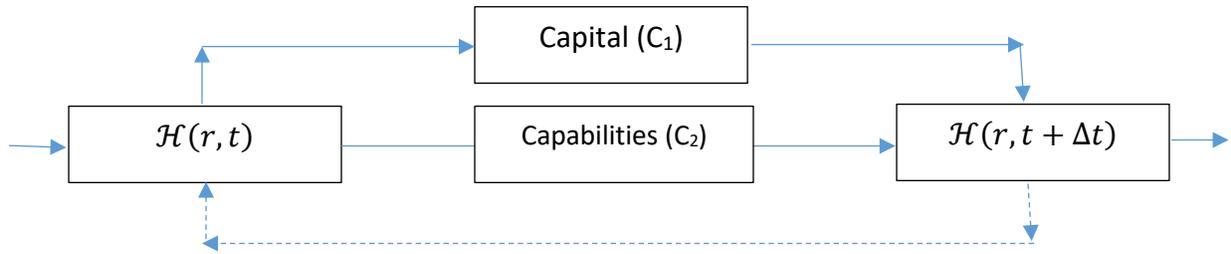

Figure 5: Feedback loop between capital and capabilities

The special conditions are:
- Normalized energy scale, $C_1 + C_2 = 1$; and k = 1;
- No explicit dependence of $\mathcal{H}$ in time, i.e $\frac{\partial}{\partial t}\mathcal{H}_s(n, t) = 0$;
- Linear or Constant body force $F_b(n, t)$ on $\mathcal{H}_s$
- Neglecting the dissipation term, i.e. $\dot{\mathbb{Q}}_s = 0$.

Under these conditions above, we may write,
$$\mathcal{H}(C_1, C_2, t) = 1 + C_1 \cdot C_2 \qquad \text{Equation 32}$$

Change of variables: y = $C_1$, x = $C_2$, $\Rightarrow \dot{y} = \frac{dC_1}{dt}$ and $\dot{x} = \frac{dC_2}{dt}$

Substituting $\mathcal{H}$ in Equation 19, the RHS terms $F_s(n, t)\,\dot{r}_n$ of Equation 28 reduces to the surface integral
$$\iint (x - y) dx\, dy = a_1 x + a_2 + \frac{1}{2}(x^2 y - xy^2)$$

The first term on the RHS of Equation 28 can be expressed as,
$$\sum_{i=1,2} \frac{\partial \mathcal{H}_s(n,t)}{\partial C_i}\,\dot{C}_1 = x\dot{y} + y\dot{x}.$$

Substituting these values in equation 28, and also including a linear body force, we can get a form of LV equation as follows:
$$x\dot{y} + y\dot{x} = a_1 x + a_2' + a_3 xy + \frac{1}{2} xy(x - y). \qquad \text{Equation 33}$$

Equation 33 is a form of LV equation. In other words, the LV type equations can be derived from the time-dependent Hamiltonian equation in the social field.

## 11.0 Implications: Physics and Economics

The social field theory claims to formalize a new form of energy, and hence it may help break the glass ceiling of utilizing thermodynamics to study social dynamics. Many scholars [64, 65, 66, 67, 68] including ET Jaynes, trusted thermodynamics to bridge the well-lit roads on each side of the natural and social science divide. A good theory can inform what there is to actually measure for a research question. This paper may provide what to look for in the sea of big data for some testable hypothesis in relation to the dynamics of human society. Ideas embedded here may provide a clue to understanding turbulence in fluid dynamics as well as the chaos in social dynamics that takes place in some parts of the world. We will address these issues in the future and as we do so we will develop further confidence in this under-explored territory of social energetics. For the time being, here are some implications for physics and economics.

   a) Physics

Is a Social Field Theory plausible? Or is it just another fantasy? Physicists are probably in the best position to evaluate these questions. An analysis of open systems may have to go beyond a narrow lens of the conservation laws. An overarching assumption such as the conservation of money takes us nowhere in our efforts to deepen our understanding of 'What is money?' The SFT is expressed here in natural units, hence it demands a "Cavendish experiment" this time in the social science domain. Such an experiment may pave the way to quantify the capabilities (PE) of an individual in a given social field. How can these ideas, if plausible, inform policy decisions, especially to help catalyze development under real circumstances on the ground?

The social field theory comes along with two fundamental postulates. One is an extension of Bohr's postulates, and the other is based on the review of the repository of knowledge developed by progenitors. These postulates need to go through rigorous experimental testing in order to prove that they can stand on their own. It is our hope that these postulates will survive because convergence is the ultimate nature of science. One measure of progress in the development of any science is its ability to make close contact with other sciences [28, 69]. We need to discover a generally acceptable conceptual framework and a language that will facilitate communication and stimulate further coherent research across the natural and social sciences beyond econophysics [70] as it exists today. Economic science may benefit most from such a framework, and ongoing synergy of physicists and economists may escalate to another level of cooperation in order to better understand the issues challenging humanity.

   b) Economics

Money is a concept of paramount significance to economic science. The energetics framework conceives of money in accordance with original insights of Howard Odum, see Chapter 4: Energy and Money [71]. Money flows in circles, but energy flows through a system and ultimately out in a degraded form. To sum up, energy and money flow hand in hand but in opposite directions. For further details, we refer readers to Howard Odum [72] and Frederick Soddy [73]; we decided not to duplicate their original insights in this short paper.

It is our hope that SFT may provide a new physically-oriented underpinning of economic science. Economics through the lens of energetics is a science up in the air, above an abstraction layer. Economic science tends to abstract many things human beings value in pecuniary terms. Such an abstraction facilitates exchanges and efficiency, but obscures a lot of dynamics important to comprehending a social system. "All non-trivial abstractions," says

Spolsky, "to some degree, are leaky." Abstraction facilitates the building up of complex systems such as the global economic system. A system based on abstraction, however, is bound to fail sooner or later to some extent. A recession may be an instance of the abstraction's failure. Let's not go too far out of our comfort zone. Economic science, we admit, is much more complex than physics because it is dealing mostly with the non-living world and active matters only very recently.

The SFT may provide a clue toward a non-overlapping definition of capital and capabilities [11]. A credit is a stake on the capabilities. Poverty relates to the energy state of an individual in a society. If an individual migrates to a new society, he can well be at another energy level than in his native society. The business cycles result from an interplay between capital and capabilities in an autonomous society which can generate a source term of its own. We witness inequality in a modern society as a possible consequence of pseudo-forces in the non-inertial social field. There are alternative physical reasons for wealth inequality [74]. A recession, like the one in 2008, may be a by-product of polarization of some sort. Equally, it could trigger a bandwagon effect influencing trust vectors among varieties of people in the global economic society. We have established here a logical explanation of why the social field is an autonomous field. The social field neither requires an external intervention to break a symmetry nor does it necessitate an 'Invisible Hand' to prescribe a function.

## 12.0   Discussion

This research is an outgrowth of our efforts to better understand the energy-poverty nexus. Energetics can pave the way for the integration of knowledge across two cultures and in time perhaps furnish an expression for the dynamics of human society. We may have come up with a viewpoint that is not visible from the natural or social science perspective alone. Equally possible, this blue-sky thinking may be considered a lax analogy by some; we admit this probable nexus portends that we go beyond our comfort zones. An academic discipline is like a valley in the mountain. One may be better off focusing on cultivating one's own valley. Nonetheless, we believe it is worth an effort. Our inspiration distills in the words of Churchill: "It is not enough that we do our best; sometimes we must do what is required." Energetics provides a map of roads over the mountains connecting these academic valleys designated as natural and social sciences. This roadmap may not be enough; we may need to pave the way for these roads at a more fundamental level. Eventually, we will also need to build and illuminate tunnels between these academic valleys for which we will rely on the further study of physicists and social scientists. This can't be another case of a fish being oblivious to water, similar to flat earth and geocentric models. We believe this is another problem we face with the utilization of human knowledge not being given to anyone in its totality [75]. The social field can't be a no man's land between the natural science and social sciences.

Social science relies on concepts based on the intuition of its forefathers who raised many important questions that are even interesting to natural scientists. The postulates based on SFT may provide deductive reasoning for some dynamics we witness in the hierarchical social field. We don't claim that the normative question in the social sciences can be inferred entirely based on scientific reasoning. Instead, SFT may provide a foundation on which some common universal principles may be extended to study both natural and social sciences. The common principle we suggest is based on an energetics that acknowledges the hierarchical structure of the social field and its underlying power dynamics. A most important element of any inquiry is the foundation on which we search for answers to many research questions. A superficial and inadequate foundation may not provide an adequate answer to a question even if we may apply the most rigorous methods at a later stage of investigation [76]. SFT may catalyze our efforts to complement our understanding of evolution and provide an alternative line of reasoning for some dynamics we witness in our society. Evolution is not just a biological process [26], it applies equally well to the human society.

Chemiosmosis [77] is a mechanism by which a biological cell extracts energy from its environment. As an aggregate of biological cells, human life can't be an exception. Human beings generate and dissipate energy in the social field, and its evolution is bounded by the same universal principles of energetics. We expand the energetic approach to connect the dots of knowledge in order to theorize human evolution at the macroscale. On the logical foundation of the SFT, we explain how the dynamics of human life are rooted in physics and thermodynamics. The same universal principles of physics manifest at scales ranging from the atomic to the human, and to the planetary scales and beyond. We propose SFT as a bridge connecting disciplines [78] such as physics and economics that may also add to the knowledge of the life sciences. Not only does SFT provide a new perspective to evolution based on the laws of physics but it also provides a logical foundation for understanding cooperative behavior [79] among human beings and human societies.

Hopefully, one day we will have a better sense of the molecular roots of evolution [80] and the principles underlying human life, as suggested by Frederick Soddy, Bertrand Russell, Manfred Eigen and many others. To fill in this gap, we must rely on something beyond a firsthand knowledge of the subject. We submit that the SFT and underlying energetics-based equations are a stopgap for our knowledge about "equations of motion for social systems". These equations may be valid in the social field but not completely valid for the social system; for the latter is made up of human-made and natural systems.

## 13.0  Conclusion

In an effort to understand human life through physics, we presented a theoretical model of a human being by the generalization of classical field theories. The social field theory formalizes the energy of an individual in the social field. On the logical foundation of the field theory, we developed a provisional version of "equations of motion for social systems" that Wolfgang Weidlich claimed non-existent in the literature. The Lotka-Volterra type equations can be derived from the time-dependent Hamiltonian equations in the social field. A model based on the social field theory that we present here may complement the existing exploratory causal mechanisms of the evolution of human society. This model may contribute to an understanding of why evolution works and open up some new perspectives. Energetics makes evolution possible. All models are wrong, but some are useful for some time during the development of our knowledge. We anticipate that an energetics model based on the social field theory be useful to bridge the terra incognito between the natural science and social science.

## 14.0   Conflicts of Interest

The authors declare that there is no conflict of interest regarding the publication of this paper.

## 15.0   Acknowledgment

We benefited from inputs of participants at multiple conferences. Especially, we would like to mention participants of the 7th BioPhysical Economics meeting (BPE-2016) [81, 82]; the 9th Biennial Conference of United States Society for Ecological Economics (USSEE); the 11th American Society of Nepalese Engineers (ASNEngr) Annual Conference 2018; and Complex Systems Group Meeting (2018) at Worcester Polytechnic Institute. We owe a long list of scholars and progenitors for their direct and indirect intellectual contributions to this paper.

We did not seek any funding for this research. RP acknowledges a sabbatical granted by the Institute of Engineering, Tribhuvan University.